\documentclass[apjl]{emulateapj}
\usepackage{natbib}
\usepackage{epsfig}
\usepackage{apjfonts}
\usepackage{subfigure}
\bibpunct{(}{)}{;}{a}{}{,}

\shorttitle{The Orbit of TWA 5A}
\shortauthors{Konopacky et al.}

\begin{document}

\title{Measuring the Mass of a Pre-Main Sequence Binary Star Through the Orbit of TWA 5A}
\author{Q.M. Konopacky\altaffilmark{1,2},
  A.M. Ghez\altaffilmark{1,2}, G. Duch\^{e}ne\altaffilmark{3}, C. McCabe\altaffilmark{4}, 
  B.A. Macintosh\altaffilmark{5}}
\altaffiltext{1}{UCLA Division of Astronomy and Astrophysics, Los Angeles, CA 90095-1547; quinn, ghez@astro.ucla.edu}
\altaffiltext{2}{Institute of Geophysics and Planetary Physics, University of
California, Los Angeles, CA 90095-1565}
\altaffiltext{3}{Laboratoire d'Astrophysique,
Observatoire de Grenoble, Universite Joseph Fourier - BP 53,
F-38041, Grenoble Cedex 9, France; Gaspard.Duchene@obs.ujf-grenoble.fr}
\altaffiltext{4}{NASA Jet Propulsion
  Laboratory, MS 183-900, 4800 Oak Grove Dr., Pasadena, CA
  91109-8099; mccabe@jpl.nasa.gov}
\altaffiltext{5}{Institute of Geophysics and Planetary
  Physics, Lawrence Livermore National Laboratory, 7000 East
  Avenue L-413, Livermore, CA 94551; bmac@igpp.ucllnl.org}

\begin{abstract}

    We present the results of a five year monitoring campaign
of the close binary TWA 5Aab in the TW Hydrae association,
using speckle and adaptive optics on the W.M. Keck 10 m
telescopes.  These measurements were taken as part of our
ongoing monitoring of pre-main sequence (PMS) binaries in an
effort to increase the number of dynamically determined PMS
masses and thereby calibrate the theoretical PMS evolutionary
tracks.  Our observations have allowed us to obtain the first
determination of this system's astrometric orbit.  We find an
orbital period of 5.94 $\pm$ 0.09 years and a semi-major axis
of 0$\farcs$066 $\pm$ 0$\farcs$005.  Combining these results
with a kinematic distance, we calculate a total mass of 0.71
$\pm$ 0.14 M$_\odot$ (D/44 pc)$^{3}$. for this system.  This
mass measurement, as well as the estimated age of this system,
are consistent to within 2$\sigma$ of all theoretical models
considered.  In this analysis, we properly account for
correlated uncertainties, and show that while these
correlations are generally ignored, they increase the formal
uncertainties by up to a factor of five and therefore are
important to incorporate.  With only a few more years of
observation, this type of measurement will allow the theoretical models to be
distinguished.  
\end{abstract}


\section{Introduction}

     Binary stars present a unique laboratory for the study of
stellar evolution, as their orbital solutions give direct mass
estimates.  Though the mass of a star is the most fundamental
parameter in determining its course of evolution, very few PMS
stars have dynamically determined masses.  In these few cases,
astrometric and spectroscopic studies have lead to PMS mass
determinations based on the orbital motion of a stellar
companion or a circumstellar disk relative to the primary star
(e.g. Ghez et al. 1995, Simon et al. 2000, Steffen et
al. 2001, Woitas et al. 2001, Tamazian et al 2002, Duchene et
al. 2003).  Once measured, these masses can subsequently be
used to constrain PMS evolutionary models.  These models have
been shown to be systematically discrepant in their
predictions by up to factors of two in mass and ten in age -
still, since there are so few well-determined PMS masses,
little can currently be done to calibrate these tracks.  This is
particularly true in the lowest mass regime, where only
three systems have total dynamical masses of less than
1M$_\odot$ and only one single star has a dynamical mass
measurement below 0.5$M_\odot$ (Hillenbrand $\&$ White 2004).
This study is part of an ongoing program to astrometrically
determine the orbits of PMS stars and to help constrain these
theoretical mass tracks.  Young stars are particularly
important to correctly calibrate, as this calibration will aid
in the subsequent calibration of young brown dwarf and
planetary models.

   The TW Hydrae association was originally discovered by
Kastner et al. (1997), with only five members confirmed at
that time.  Since its discovery, 25 total members have been
identified (e.g., Song et al. 2003, Makarov $\&$ Fabricius
2001).  The association has been shown to be quite young
($\thicksim$8-12 Myr) via lithium abundance tests, space
motions, and placement of its members on the HR diagram (e.g.,
Zuckerman $\&$ Song 2004).  Additionally, TW Hydrae is quite
nearby, with an average distance of $\thicksim$50 pc (e.g.,
Makarov $\&$ Fabricius 2001, Mamajek 2005).  This makes TW
Hydrae one of the closest associations of young stars to the
Earth and thus is an ideal region for studying spatially
resolved PMS binaries, as they are likely to have orbital
periods as short as a few years.

   With these ideas in mind, we began to monitor TWA 5, the
fifth of the five original members of the TW Hydrae
association identified by Kastner et al. (1997).  TWA 5 is
composed of at least three components.  The pair that we
analyze here is TWA 5Aa-Ab, which had a separation of
$\thicksim$0$\farcs$06 when it was discovered by Macintosh et
al. (2001).  TWA 5A also has a brown dwarf companion, TWA 5B,
located $\thicksim$2$\arcsec$ away (Webb et al. 1999, Lowrance
et al. 1999, Neuhauser et al. 2000, Mohanty et al. 2003).
Finally, TWA 5Aab is suspected to contain at least one
spectroscopic pair based upon large radial velocity variations
(Torres et al. 2003, Torres 2005 private communication).

       In this paper, we present the results of six years
of speckle and adaptive optics (AO) observations of TWA 5Aab
with the W.M. Keck 10 m telescopes.  In $\S$2, we describe our
data reduction techniques, and in $\S$3 we present our orbital
solution for the system.  In $\S$4, we discuss comparisons of
our derived orbital parameters with mass and age predictions
from theoretical PMS tracks and make recommendations
for future studies of this system.
 
\section{Observations and Data Analysis}
\subsection{Speckle Data}

\begin{figure*}
\epsscale{0.75}
\plotone{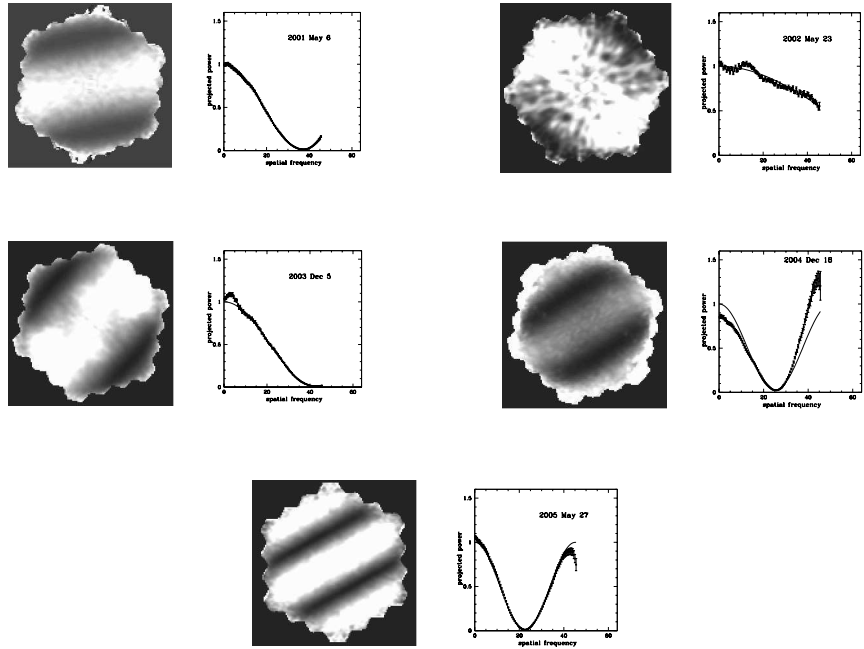}
\caption{The two-dimensional visibilities for all the speckle
  data are shown in combination with plots of one-dimensional cuts through the
  visibilities.  The points represent actual data, while the
  solid lines show the best fit of Equation 1 to this data.
  The data from 2002 clearly illustrates that the separation
  of the binary during this epoch was below the diffraction
  limit.  The degeneracy of the separation and flux ratio in
  this fit (where the first minimum was not reached)
  necessitated the fixing of the flux ratio in this epoch to
  obtain the correct separation.  The results of this analysis
  can be seen in Figure 3.  The slight discrepancy in
  the 2004 fit comes from discrepancy between the calibrator
  and the object, stemming from large scale changes in seeing
  on short timescales.  However, the large error bars on this
  data point account for this variation and thus they do not
  weigh heavily in our orbital fit.}
\end{figure*}

\begin{deluxetable*}{lcclllcc}
\tabletypesize{\scriptsize}
\tablecolumns{8}
\tablewidth{0pc}
\tablecaption{TWA 5A Binary Star Parameters}
\tablehead{
\colhead{Date of} &  \colhead{Filter} & \colhead{$\lambda_{o}$} & \colhead{Separation} & \colhead{Position Angle} &
\colhead{Flux Ratio} & \colhead{Speckle} &
\colhead{Source\tablenotemark{a}} \\
\colhead{Observation} & \colhead{} & \colhead{($\mu$m)} & 
\colhead{(arcseconds)} & \colhead{(degrees)} &
\colhead{(Aa/Ab)} & \colhead{or AO?} & \colhead{}
}
\startdata
2000 Feb 20 & H & 1.648 & 0.0548 $\pm$ 0.0005  & 205.9 $\pm$ 1.0 & 1.09
$\pm$ 0.02   & AO & 2 \\
2000 Feb 22 & K' & 2.127 & 0.054 $\pm$ 0.003 & 204.2 $\pm$ 3.0 & 1.11
$\pm$ 0.07  & AO & 3 \\
            & H & 1.648 & & & 1.09 $\pm$ 0.08 & AO & 3 \\
            & J &  1.26 & & & 0.94 $\pm$ 0.05 & AO & 3 \\
2001 May 06 & K & 2.2 & 0.0351 $\pm$ 0.0002  & 12.67 $\pm$ 1.06 & 1.26 $\pm$
0.09 & Sp & 1 \\
2002 May 23 & K & 2.2 & 0.013 $\pm$ 0.003 & 313.66 $\pm$ 2.99 & 1.24
$\pm$ 0.08\tablenotemark{b} & Sp & 1 \\
2003 Dec 05 & K & 2.2 & 0.0306 $\pm$ 0.0004  &  227.41 $\pm$ 5.49 &  1.22 $\pm$
0.04 & Sp & 1 \\
2004 Dec 18 & K & 2.2 & 0.0515 $\pm$ 0.0009 & 32.10 $\pm$ 2.22  & 1.39 $\pm$
0.09 & Sp & 1 \\
2005 Feb 16 & FeII & 1.65 & 0.053 $\pm$ 0.001 & 32.59 $\pm$ 5.22 &
1.29 $\pm$ 0.18 & AO & 1 \\
2005 May 27 & K & 2.2 & 0.0574 $\pm$ 0.0003 & 29.68 $\pm$ 0.35 & 1.23
$\pm$ 0.04 & Sp & 1 \\
2005 Dec 12 & Kp & 2.124 & 0.0571 $\pm$ 0.001 & 29.99 $\pm$ 2.26 & 1.10
$\pm$ 0.05 & AO & 1 \\
            & H & 1.633 & 0.0571 $\pm$ 0.002  &  28.89 $\pm$ 0.99 &
1.09 $\pm$ 0.03 & AO & 1 \\
            & J & 1.248 & 0.0568 $\pm$ 0.005 & 28.41 $\pm$ 3.15 & 1.05
$\pm$ 0.13 & AO & 1 \\
\enddata
\tablenotetext{a}{1 = This work; 2 = Macintosh et al. 2001; 3
  = Brandeker et al. 2003}
\tablenotetext{b}{Flux ratio fixed in this epoch}
\end{deluxetable*}

     TWA 5Aab was observed using the Keck I 10 m telescope
 with the facility Near Infrared Camera (NIRC, Matthews $\&$
 Soifer 1994, Matthews et al. 1996) roughly once a year from
 2001 to 2005 (see Table 1 for exact dates). In its high
 angular resolution mode, NIRC has a pixel scale of 
 20.44 $\pm$ 0.03 mas/pixel (see Appendix A for details of NIRC's pixel
 scale and orientation).  Three to four stacks
 of 190 images, each 0.137 seconds long, were obtained through
 the K band-pass filter ($\lambda_o$ = 2.2 $\mu$m,
 $\Delta\lambda$ = 0.4 $\mu$m).  They were taken in stationary
 mode, meaning the Keck pupil is fixed with respect to the
 detector during the observations, but the sky rotates.  The
 rotation rate is sufficiently slow that it is negligible for
 individual exposures, but not over the entire stack.  Stacks
 of four dark frames were also taken with each object stack.
 Additionally, identical stacks of a point source calibrator
 star (TWA 1 for 2001-2002 observations, TWA 7 for 2003-2005
 observations) and an empty portion of sky were obtained
 immediately before or after the target stacks.

     The stacks are processed using image reduction and
speckle data anaylsis techniques.  Specifically, each image is
first dark and sky subtracted, flat fielded, and bad
pixels repaired.  The images are then individually
corrected for a minor optical distortion in the NIRC
camera (J. Lu et al. in prep).
The object, calibrator, and sky stacks are then
Fourier transformed and squared to obtain stacks of power
spectra.  Next, the calibrator stacks and sky stacks are each
averaged together (without rotation).  The extraction of the
object's intrinsic power spectra utilizes the convolution theorem, which
gives the relation:
\begin{equation}
O = \frac {I - <S>}{<C> - <S>}  ,
\end{equation}
where O is the object power spectrum, I is the initial squared
  Fourier transform of the object, $<$S$>$ is the averaged, squared
  power spectrum of the sky, and $<$C$>$ is the averaged, squared
  power spectrum of the calibrator.  Each power spectrum of the object
  is then rotated so that north is up; these
   rotated power spectra are combined to
  obtain an average power spectrum for the stack. 
   Finally, those individual stacks were also averaged together to produce a
final power spectrum.  This sequence is necessary due to the
  lack of azimuthal symmetry of the Keck I pupil.

     Since a binary star in the image domain is essentially
two delta functions, the power spectra can be approximated as
a sinusoid with the functional form:
\begin{equation}
P(\vec f)= \frac {R^2 + 1 + 2R cos(2\pi \vec f \cdot \vec s)}{R^2 + 1 + 2R} ,
\end{equation}
where R is the flux ratio of the binary star and $\vec s$ is
the vector separation of the two stars.  This function is fit
to our two-dimensional power spectra over all spatial
frequencies from 2.68 arcsec$^{-1}$ to 17.6 arcsec$^{-1}$
(this procedure is described in detail in Ghez et al. 1995).
The lower cutoff is imposed to avoid spatial frequencies that
are corrupted by small changes in the atmospheric conditions
between the object and the calibrator, whereas the upper
cutoff is imposed to reject excess noise at the highest
spatial frequencies.  This fitting procedure gives a very
accurate estimate of the binary star's separation and flux
ratio for components separated by more than $\lambda$/2D,
leaving only a 180$^o$ position angle ambiguity.  This
ambiguity is resolved by reanalyzing the original images using
the method of shift and add.  The brightest speckle in the
speckle cloud in each image of the stack is shifted to a
common position, and then all the images are added together to
produce a diffration-limited core surrounded by a large,
diffuse halo.  These images allow us to determine the correct
orientation of the position angles of TWA 5A.  In one epoch
(2002), the binary separation was less than $\lambda$/2D, and
hence the first minimum of the power spectrum was not
measured; without this measurement, the separation and the
flux ratio are degenerate.  For this epoch, we fix the flux
ratio to the weighted average of the K-band flux ratio
measurements of all the other epochs and fit only the
separation, as there is no evidence of statistically
significant photometric variability
over the course of our observations.  Uncertainties for all
parameters are determined by fitting each of the individual,
stack-averaged images that contribute to the final image by
the same procedure, and then taking the RMS of those values
with respect to the average value; for the 2002 epoch we also
account for the uncertainty in the weighted-average flux
ratio, which is taken to be the RMS of the individual K-band
flux ratio measurements at all other epochs.

\subsection{Adaptive Optics Data}

\begin{figure}
\epsscale{0.7}
\plotone{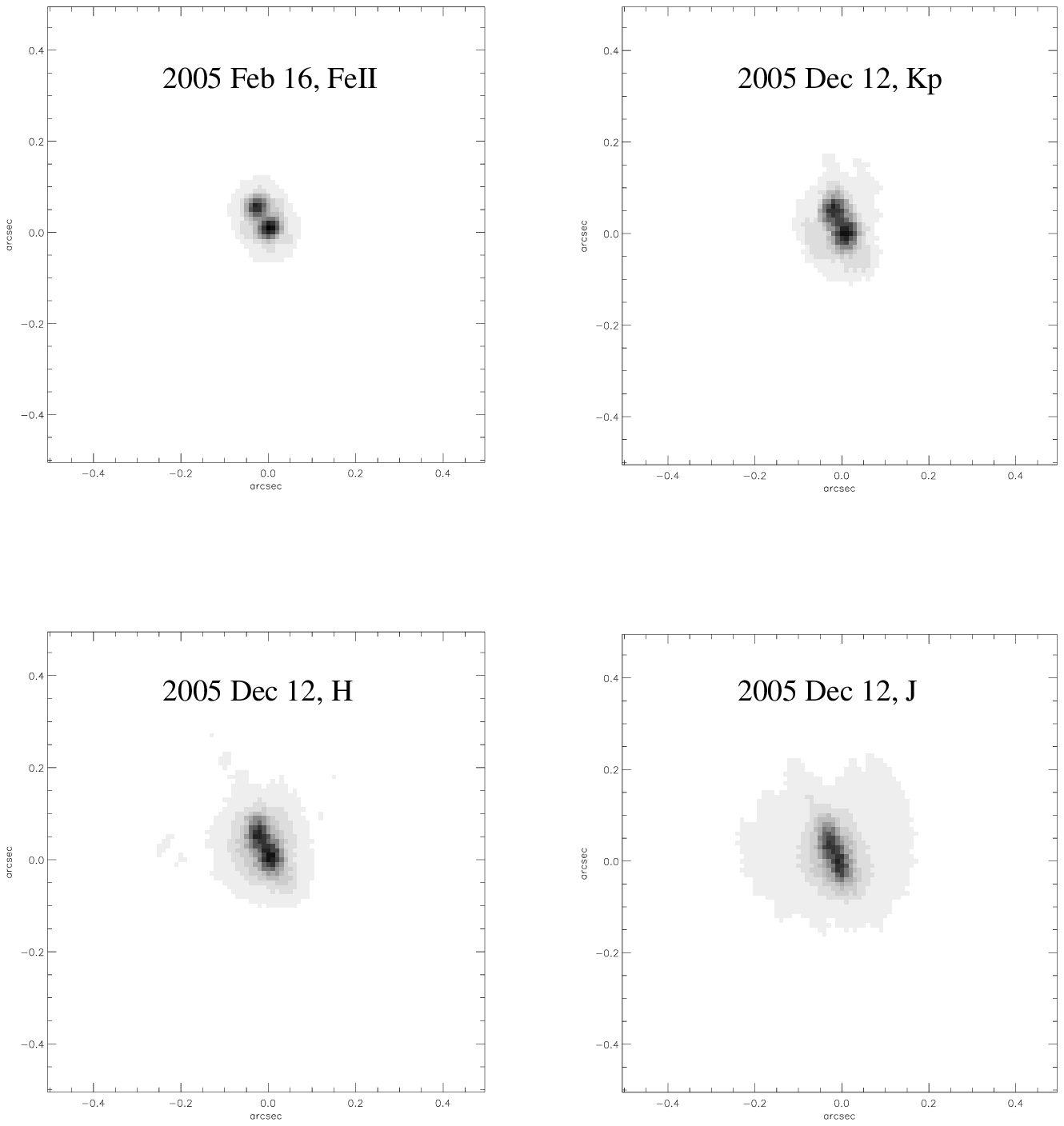}
\caption{NIRC 2 AO images of TWA 5A taken on 2005 February 16
  and 2005 December 12.  In all images, north is up and east
  is to the left.  Component TWA 5Aa is to the southwest and
  component TWA 5Ab is to the northeast.}
\end{figure}

TWA 5Aab was also observed using the Keck II 10 m telescope
with the AO system (Wizinowich et al. 2000) and the facility near-infrared camera, NIRC
2 (K. Matthews et al., in prep), on 2005 February 16 and again on
2005 December 12.  For these measurements, we used
observations of the Galactic Center to establish that NIRC 2's
narrow camera (which we used) has a plate scale of 9.961 $\pm$
0.007 mas pixel$^{-1}$ and columns that are at a PA of -0.015
$\pm$ 0.134$^o$ relative to North (J. Lu et al., in prep).  In
February, five images, each of 0.2 second exposure time and 30
coadds, were taken through the FeII narrow band pass filter
($\lambda_o$ = 1.65 $\mu$m, $\Delta\lambda$ = 0.03 $\mu$m).
In December, six, six, and three images of 0.181 second
exposure time and 50 coadds were taken through the J
($\lambda_o$ = 1.248 $\mu$m, $\Delta\lambda$ = 0.163 $\mu$m),
H ($\lambda_o$ = 1.633 $\mu$m, $\Delta\lambda$ = 0.296
$\mu$m), and K-prime ($\lambda_o$ = 2.124 $\mu$m,
$\Delta\lambda$ = 0.351 $\mu$m) band pass filters,
respectively.  These images were processed using the same
standard data reduction techniques listed above, and then
shifted and combined to produce a final image.

Astrometry and flux ratios were then obtained using the IDL
package StarFinder (Diolaiti et al. 2000).  The wide brown
dwarf companion to the system, TWA 5B, was visible in all AO
images taken and thus was used as the empirical point-spread
function required by the StarFinder fitting algorithm.  Both
components of the close binary were successfully fit by
StarFinder in all cases.  Errors were calculated by fitting
the components in all individual images that contributed to
the combined images and finding the RMS of the values derived
therein.
  
\section{Results}

   Figure 1 shows the calibrated power spectra and our
resulting fits for all five speckle measurements of TWA 5Aab
and Figure 2 displays our 2005 AO images.  We supplement our
observations with the following measurements that also
spatially resolved TWA 5Aab: the original discovery
measurement by Macintosh et al. (2001), and another taken two
days later by Brandeker et al. (2003).  Table 1 lists all
separation, position angle, and flux ratio measurements that
are used in this study.

    Over the six years that the components of TWA 5A have
been spatially resolved, the binary has undergone a
full orbit (see Figure 3), allowing an accurate estimate of its
orbital parameters.  We calculate an orbital solution for TWA
5A using the Thiele-Innes method (e.g., Hilditch 2001),
minimizing the $\chi^2$ between the model and the
measurements, which are converted from angular separation and
position angle to right ascension and declination.  Our model
incorporates the following 7 standard orbital elements: P
(period), A (semi-major axis), e (eccentricity), i
(inclination), T$_o$ (time of periapse passage), $\Omega$
(longitude of the ascending node), and $\omega$ (argument of
pericenter).  With nine two-dimensional astrometric
measurements, there are eleven degrees of freedom in our fit.
The best fitting model produced a $\chi^2$ of 8.91 with 11
degrees of freedom.  Furthermore, of the nine data points used
in the fit, seven are within 1$\sigma$ and two are within
2$\sigma$ of the model, suggesting that our fit is good.  The
1$\sigma$ uncertainties in the model parameters are estimated
by changing the values of $\chi^2$ by one (Bevington $\&$
Robinson 1992).  Table 2 lists the best-fit orbital parameters
and their uncertainties and Figure 3 shows our solution for
the projected orbit of TWA 5A.  This astrometric solution
yields a mass 0.71 $\pm$ (0.14 $\pm$ 0.19) M$_\odot$, where the two
sources of uncertainty stem from our orbital
solution and the from the 9$\%$ uncertainty in the
kinematic distance estimate to TWA 5A (Mamajek 2005). 

\begin{deluxetable}{lc}
\tabletypesize{\scriptsize}
\tablecolumns{2}
\tablewidth{0pc}
\tablecaption{Orbital Parameters of TWA 5A}
\tablehead{
}
\startdata
P (years) & 5.94 $\pm$ 0.09 \\
A ($\arcsec$) & 0$\farcs$066 $\pm$ 0.005 \\
T$_o$ (years) & 2004.34 $\pm$ 0.09 \\
e & 0.78 $\pm$ 0.05 \\
i (degrees) & 97.4 $\pm$ 1.1 \\
$\Omega$ (degrees) & 37.4 $\pm$ 0.9 \\
$\omega$ (degrees) & 255 $\pm$ 3 \\
\enddata
\tablecomments{We note that the value of $\Omega$ is actually
  subject to a 180 degree ambiguity without three dimensional
  velocity information.}
\end{deluxetable}

\begin{figure}
\epsscale{1.0}
\plotone{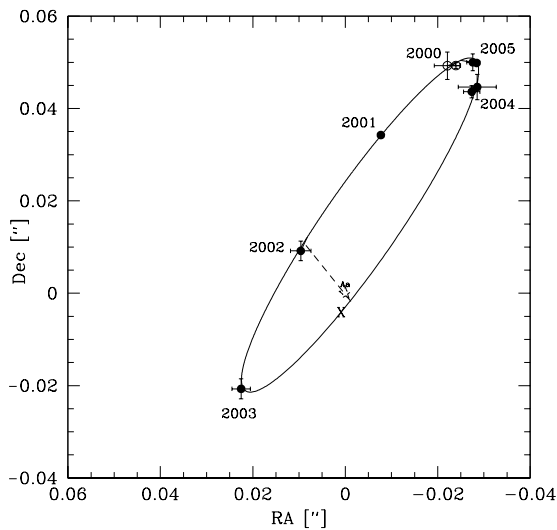}
\caption{Orbital solution for TWA 5Aa and TWA 5Ab.  Filled
  circles represent the data taken in this study, while the
  open circles represent data taken from the literature (both
  in 2000).  The star at [0,0] marks the position of TWA 5Aa;
  the lines from the data points to the ellipse indicate where
  the fit believes the point should lie on the orbit.  The
  dashed line through the center represents the line of nodes
  and the 'X' marks the location of closest approach.  The
  parameters for this orbit are given in Table 2.}
\end{figure}

\section{Discussion and Conclusions}

\begin{deluxetable*}{lcccccccc}
\tabletypesize{\scriptsize}
\tablecolumns{9}
\tablewidth{0pc}
\tablecaption{TWA 5 System Photometry}
\tablehead{
\colhead{Component} & \colhead{Sp. Ty.} & \colhead{m$_K$} & \colhead{m$_H$}
& \colhead{m$_J$} & \colhead{J-H} & \colhead{H-K} &
\colhead{Log[Temp] (K)} & \colhead{Log[Lum] (L$_\odot$)}
}
\startdata
TWA 5Aa & M1.5 $\pm$ 0.5 & 7.39 $\pm$ 0.04 & 7.69 $\pm$ 0.04 & 8.40 $\pm$
0.07 &  0.71 $\pm$ 0.08 & 0.30 $\pm$ 0.06 & 3.56 $\pm$ 0.01 & -0.80 $\pm$ 0.08 \\
TWA 5Ab & [M1.5 $\pm$ 0.5] & 7.62 $\pm$ 0.08 & 7.79 $\pm$ 0.05 & 8.45
$\pm$ 0.15 &  0.66 $\pm$ 0.16 & 0.17 $\pm$ 0.09 & 3.56 $\pm$ 0.01 & -0.84 $\pm$ 0.08 \\
\enddata
\tablecomments{The apparent K, H, and J band magnitudes for TWA5Aa
  and Ab are
  computed using 2MASS measurements of the combined
  magnitudes and the flux ratio measurements given in Table
  1.  For the H and J band, we use the most recent AO flux
  ratio measurements taken with NIRC 2 in 2006 December.  For
  the K band, we use the average flux ratio from all speckle
  measurements (excluding the unresolved measurement in 2002). 
The spectral type for TWA 5Ab is assumed to be roughly the
same as TWA 5Aa, given that the two components have nearly
equal brightness.
Temperatures are estimated from the scalings given in Luhman
et al. 2003, while bolometric luminosities are calculated
using the kinematic distance and H band bolometric corrections
(Luhman 2005, private communication).}
\end{deluxetable*}

    With a well-determined astrometric orbit for TWA 5A, it is
possible to begin to compare dynamical estimates of mass with
those inferred from theoretical PMS tracks.  By comparing the
quantity M/ (D/44 pc)$^{3}$, we preserve the high precision of
the astrometric data in the analysis.  Here we investigate the
models by Baraffe et al. 1998 ($\alpha$ = 1.0), D'Antona $\&$
Mazzitelli 1997, Palla $\&$ Stahler 1999, Siess et al. 2000,
and Swenson et al. 1994.

Model estimates of mass and age require both effective
temperature and bolometric luminosity as inputs.  Effective
temperatures are estimated from the unresolved spectral type
for TWA 5A and component flux ratios.  The photometric
analysis performed here and elsewhere (see Table 1) shows that the two
components of TWA 5A have nearly equal brightness out to 1.6
$\mu$m.  We therefore assume that the components have the same
spectral type and assign it to be M1.5 $\pm$ 0.5, the spectral
type found from a spatially unresolved spectrum (Webb et
al. 1999).  This is consistent with the J-H colors for each
component, which are calculated by combining the 2MASS
unresolved magnitudes with the flux ratios measured here
(J-H[M1.5] = 0.67; Leggett et al. 1992).  We estimate the
temperature of these components using a conversion of spectral
type to effective temperature given in Luhman et al. (2003).
These temperatures are intermediate between dwarf and giant
stars, thus making them appropriate for PMS stars like TWA 5A.
Bolometric luminosities are estimated using our H band
magnitudes and the corresponding H band bolometric corrections
for PMS stars (Luhman 2005, private communication), along with
the kinematic distance for TWA 5A of 44 $\pm$ 4 pc (Mamajek
2005).  These input values are given in Table 3.

The uncertainties in M/ (D/44 pc)$^{3}$ and age from each
model are estimated through Monte Carlo simulations.  Our
Monte Carlo simulation was run with $10^5$ points sampled from
random, independent gaussian distributions of H-band flux
densities and flux density ratios, temperatures for each
component, and distance.  This allowed us to correctly account
for the correlation between input values.  In particular, the
bolometric correction used for determining the luminosity
stems from the effective temperature.  Additionally, the
luminosities of the primary and the secondary are correlated,
as they are calculated from the same parameters, namely the
total flux and the flux ratios.  Each of of these values are
converted into sets of temperatures and bolometric
luminosities, which are is fed into the models to produce
estimates of masses and ages for each component.  Using the
distance assumed for each run, we convert the total mass of
both components to M/ (D/44 pc)$^{3}$.  We emphasize that
though the predictions of the models are clearly not
independent of the distance uncertainty, by using this
quantity as opposed to the total mass, we can leave the
distance uncertainty out of our dynamical measurement.  Thus,
we then generate contours that correctly represent the
1$\sigma$ uncertainties on the masses and ages derived from
the model tracks.

Figure 4 shows the dynamical mass normalized by the
distance-cubed and with it the estimated age of the TW Hydrae
association (10 $\pm$ 2 Myr, based on comparison with the
$\beta$ Pictoris moving group, Zuckerman $\&$ Song 2004),
compared with the model predictions as determined by the Monte
Carlo simulations.  While the input values are generally
treated as independent variables, the Monte Carlo simulations
demonstrate that it is important to consider the existing
correlations between input values, as the resulting
uncertainties are otherwise significantly underestimated (by
up to a factor of five depending upon the model).  With this
proper treatment, the dynamical mass and age estimates are
within 2$\sigma$ of all five tracks.  

\begin{figure}
\epsscale{1.0}
\plotone{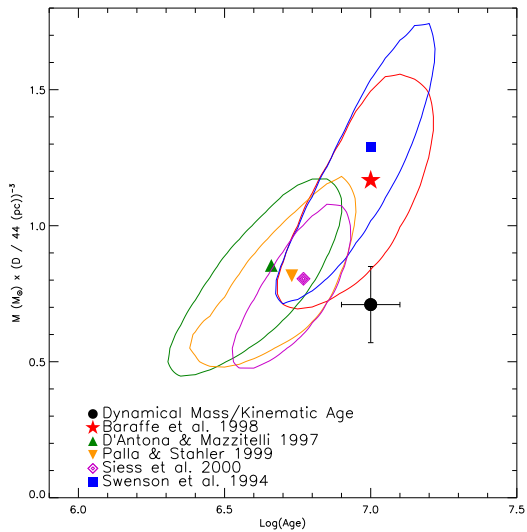}
\caption{Mass and age for TWA5Aa+b calculated from the
  dynamical solution and predicted from each of the five
  theoretical models considered.  The age shown here is the
  age predicted for the primary, but the age predictions for
  both components are consistent with each other.  The masses
  are divided by the distance cubed.  Though the predictions
  of the models do include the distance uncertainty, by using
  the quantity M/ (D/44 pc)$^{3}$, we insure that only the model
  contours are affected by distance uncertainty, avoiding
  correlation between the models and our dynamical measurement.}
\end{figure}

While it is reassuring that current models agree with these
measurements, we are unable to distinguish between the model
tracks.  Future improvements in the precision of both the
distance and the temperature would allow for a more
illuminating comparison between the theoretical models.  The
temperature uncertainties on the components of TWA 5A could be
substantially reduced with spatially resolved spectral types,
which should drive the uncertainty in the temperature down to
$\pm$ one quarter of a spectral subclass.  Uncertainties in
the distance could also be substantially reduced.  Typical
uncertainties in new parallax measurements are on the order of
a few milliarcseconds, with improvements on these results
promised in the near future (Vrba et al. 2004).  A separate
method to determine the distance is the addition of spatially
resolved radial velocities, which would allow an independent
distance estimate from the orbital parameters.  In principle,
a factor of two improvement in the distance uncertainty could
be achieved.  These radial velocity measurements would also
improve the total mass estimate, as would additional
astrometric measurements, and would also eventually allow for
the determination of \textit{individual} masses of each of the
components.  However a potential complication is the possible
existence of an additional component in the system (Torres et
al. 2003).  Another possible method of determining the
individual masses of the components of TWA 5A is to use
absolute astrometry with respect to TWA 5B, as was done
recently for T Tau S (Duch\^{e}ne et al. 2006).  If we just
take the improvements expected from future parallax
measurements, spatially resolved spectra, and additional
astrometry, we expect to be able to distinguish between the
tracks at the $\thicksim$3$\sigma$ level.

Future spatially resolved spectral types would also address
the marginal inconsistency between the J-H and H-K colors for
the primary.  Currently, we have assumed that this arises from
a small K-band excess (2$\sigma$).  Torres et al. (2003)
suggest that an additional component may be present in the
system.  If this component were particularly low mass, it
could give rise to the apparent K band excess of the primary.
Alternatively, the infrared excess could arise from
circumstellar material.  Mohanty et al. (2003) report the
detection of strong H$\alpha$ emission in an unresolved, high
resolution spectrum of TWA 5A, implying at least one of the
components is accreting.  However, no mid-infrared excess has
been detected in unresolved measurements.  Given the tightness
of the TWA 5A binary, any disk material is likely to be
localized to a very radially thin reservoir of material near
the dust sublimation radius.  In either case, disk or low mass
companion, the cause of the infrared excess should not have a
large impact on the track comparison, as the comparison was
performed at H-band, where the luminosities do not appear to
be significantly effected by this excess.

   In summary, our solution to the orbit of this system and the subsequent
   determination of its mass shows that these young,
   nearby associations of stars are excellent laboratories for
   the study of low-mass star formation.  There are likely
   other systems in TW Hydrae with similar close companions
   that will yield more mass estimates within a short time,
   much like we have seen here.  Thus monitoring of these
   systems will greatly aid in the constraint of PMS mass
   tracks in the near future.

\acknowledgements

The authors thank observing assistants Joel Aycock, Gary
Puniwai, Madeline Reed, Gabrelle Saurage, and Terry Stickel
for their help in obtaining the observations and Fabio
Altenbach, Randy Campbell, Jessica Lu, Kevin Luhman, Willie
Torres, and Ben Zuckerman for helpful discussions.  We also
thank Keivan Stassun for constructive feedback as referee.
Support for this work was provided by the NASA Astrobiology
Institute, the NSF Science \& Technology Center for AO,
managed by UCSC (AST-9876783), and the Packard Foundation.
Portions of this work were performed under the auspices of the
U.S. Department of Energy by the University of California,
Lawrence Livermore National Laboratory under contract
No. W-7405-Eng-48.  This publication makes use of data
products from the Two Micron All Sky Survey, which is a joint
project of the University of Massachusetts and the Infrared
Processing and Analysis Center/California Institute of
Technology, funded by the National Aeronautics and Space
Administration and the National Science Foundation.  The
W.M. Keck Observatory is operated as a scientific partnership
among the California Institute of Technology, the University
of California and the National Aeronautics and Space
Administration. The Observatory was made possible by the
generous financial support of the W.M. Keck Foundation.  The
authors also wish to recognize and acknowledge the very
significant cultural role and reverence that the summit of
Mauna Kea has always had within the indigenous Hawaiian
community.  We are most fortunate to have the opportunity to
conduct observations from this mountain.

\appendix

\section{NIRC Plate Scale and Orientation}

\begin{deluxetable}{lcc}
\tabletypesize{\scriptsize}
\tablecolumns{3}
\tablewidth{0pc}
\tablecaption{Absolute NIRC Position Angle Offsets and Uncertainties}
\tablehead{
\colhead{Epoch(s)} & \colhead{No. of} &
\colhead{Absolute} \\
\colhead{} & \colhead{Measurements} &
\colhead{PA (deg)}
}
\startdata
1998 April - 1998 August & 4 & -0.40 $\pm$ 0.135 \\
1998 October - 2002 July & 11 & -0.884 $\pm$ 0.143 \\
2003 April - 2003 September  & 3 & 0.761 $\pm$ 0.135 \\
2004 April - 2005 May & 4 & -0.728 $\pm$ 0.196 \\
\enddata
\end{deluxetable}

NIRC's pixel scale and orientation are calibrated relative to
NIRC 2 using observations of the Galactic center (see Table
5).  As discussed in section $\S$2.2, NIRC 2's absolute offset
with respect to north has been accurately calibrated, so that
by calibrating NIRC with respect to NIRC 2, we can use a
simple coordinate transform to get its absolute orientation as
well.  While NIRC's pixel scale has been stable, known
engineering adjustments have introducted slight rotations in
the camera over time.  Four out of five of TWA5A speckle
measurements were taken at a time with NIRC's
orientation is well-characterized by the Galactic center
experiment.  The 2003 measurement, however, was taken during a
period of multiple engineering adjustments and no Galactic
center data.  We therefore bound its orientation by the
   measurements taken of the Galactic Center just before and
   after it, which leads to an absolute offset of 0.032
   $\pm$ 0.719$^o$.  For all of our observations, we use a
   constant plate scale of 20.45 $\pm$ 0.03 mas/pixel.

\end{document}